\documentstyle[prl,aps]{revtex}

\draft

\input epsf         

\begin{document}

\twocolumn


\title{Spectroscopic electric-field 
measurement in laser-excited plasmas}

\author{S. K. Dutta, D. Feldbaum, G. Raithel}

\address{University of Michigan, Physics 
Department, Ann Arbor, MI 48109-1120}

\date{\today}
\maketitle

\begin{abstract}
The electric field in mm-sized
clouds of cold Rb$^{+}$-ions is measured 
using the Stark effect of Ryd\-berg atoms embedded in the 
ion clouds.  The ion clouds  
are produced by UV photoionization of laser-cooled rubidium
atoms in a magneto-optic trap. 
In very dense ion clouds, 
the Ryd\-berg atom excitation spectra reflect the presence
of a neutral plasma core and a 
positively charged ion shell.   
\end{abstract}



Highly excited Ryd\-berg atoms \cite{Gal} 
are very sensitive to static and AC electric
and magnetic fields, and can be
used for the spectroscopic detection and 
compensation of such fields. 
Ryd\-berg atoms have been 
used to detect microwave and FIR radiation \cite{FIR},
to measure and to compensate  weak static electric 
fields  \cite{Neu87}, 
and to study  QED level shifts and cavity-QED effects 
involving only few microwave photons \cite{QED}.
The spectroscopy of Ryd\-berg atoms 
has been employed to diagnose electric fields in 
DC and high-frequency discharge plasmas \cite{Gan86}.
In this paper, we use Ryd\-berg atom spectroscopy to 
determine the electric field in millimeter-sized,  
strongly coupled plasmas  that since
recently can be generated by the laser
excitation of laser-cooled clouds of atoms \cite{Rol99}.
These plasmas hold the promise of 
bearing novel low-temperature recombination \cite{Hahn97}
and oscillation \cite{osc} behavior.
Further, there is evidence 
for a Mott transition, by which 
strongly-coupled plasmas and/or dense gases of
Ryd\-berg atoms may form a new state of highly excited, 
metastable matter \cite{RM}. To investigate
these and other issues, one could utilize the non-invasive
field-measurement method employed in the present paper.

Our experiment has recently been described
\cite{zeke}. In a two-step optical excitation, 
laser-cooled rubidium atoms 
confined in a volume $<1$mm$^{3}$ are first excited 
from the ground state $5S_{1/2}$ to
the $5P_{3/2}$ state, which has an ionization wavelength 
of $\lambda_{\rm ion} = 479.1$nm.
A UV laser pulse ($\lambda_{\rm UV}=355$nm, duration$<$10ns, pulse
energy up to 10mJ) then partially ionizes the $5P_{3/2}$-atoms,
producing photoelectrons with 0.9eV initial 
kinetic energy. The photoelectrons
escape on a time scale of a few ns,
leaving behind a cloud
of slow Rb$^{+}$-ions. About 23ns after the UV pulse and 
while the lower-step laser is still on, 
the remaining $5P_{3/2}$-atoms that are 
embedded in the ion plasma
are excited by a blue dye laser pulse 
($\lambda \approx 480$nm,  duration$<$10ns, bandwidth 
$\approx 10$GHz). The ion cloud does not
significantly expand before the blue pulse arrives. The blue
pulse excites bound Ryd\-berg states, the Stark effect 
of which we use to measure the plasma electric fields.

We detect the Ryd\-berg atoms using 
the mechanism explained in the following. The
ion plasma represents an electron trap with a depth $U_{0}$ 
given by the ion number, the 
cloud size and the profile of the charge 
distribution \cite{Rol99}.  In most of our
experiments $U_{0}<0.9$eV, in which case
no or only few UV-generated 
photoelectrons are retained in the
electron trap (the ones that are moderated 
to low enough energy on their way out).
However, the plasma electron trap becomes 
populated with  electrons of about $10$meV energy 
due to direct photoionization 
of $5P_{3/2}$-atoms by the ASE 
with $\lambda < \lambda_{\rm ion}$ contained in 
the blue laser pulse and due to 
thermal ionization of a fraction 
of the bound Ryd\-berg atoms excited by the blue pulse \cite{zeke}.
Within a few $\mu$s,
$l$-changing collisions between the trapped electrons and the
Ryd\-berg atoms efficiently promote the latter 
into long-lived high-$\langle l \rangle$ states. During the subsequent Coulomb expansion 
of the plasma, which typically takes a few tens of $\mu$s,
the depth $U_{0}$ of the plasma electron trap 
approaches zero. 
During that time, the electrons trapped in the plasma
are gradually released, producing a plasma decay
signal in our microchannel-plate (MCP) electron detector,
which is located about $10$cm from the atomic cloud.
A large fraction of the surviving high-$l$ Ryd\-berg atoms
thermally ionize on a time scale of order $10$ms \cite{zeke},
which exceeds the plasma decay time
by about a factor of thousand.
The delayed thermal electrons originating in the
high-$l$ Ryd\-berg atoms are counted in a gate 
that starts well after the plasma decay, yielding
a noise-free signal of the Ryd\-berg atom excitation. 
Ryd\-berg excitation spectra are obtained by recording 
the delayed electron counts as a function of the
wavelength of the blue laser.

\begin{figure}  [h]
\centerline{\ \epsfxsize=3.5in \epsfbox{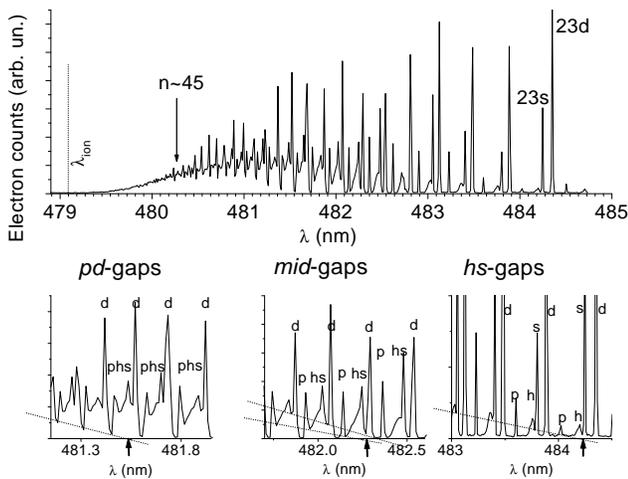}}
\caption
{Experimental Ryd\-berg excitation spectrum of 
$^{87}$Rb 5P$_{3/2}$ atoms embedded
in a laser-generated ion plasma.
The three types of spectral 
ranges with low oscillator strength,
denoted $hs$-, $mid$- and $pd$-gaps,  become filled in
at certain critical wavelengths that depend 
on the strength of the plasma electric field.
The detail plots in the lower row show how we
determine the critical wavelengths.}
\label{spectra1}
\end{figure}

Fig.~\ref{spectra1} shows a typical experimental 
Ryd\-berg excitation spectrum of atoms
in an electric-field-bearing plasma.
The field manifests itself through
a) the appearance of $p$- and 
hydrogenic lines, labeled $p$ and $h$ in Fig.~\ref{spectra1}, 
b) termination of the discrete Ryd\-berg series at a certain
field-dependent value of the principal quantum number $n$, 
and c) continuum lowering.
While the effects a) and b) are
solely determined by the state of the ion plasma
at the time instant of the Ryd\-berg excitation, the
feature c) also depends on the 
history of the Ryd\-berg atoms 
after their excitation. 
Since we intend to analyze the plasma 
at the time instant of the Ryd\-berg atom excitation, we 
disregard the feature c) as a field measurement tool.
We also don't use b), because
we have found that in many of our spectra the laser linewidth
terminates  the Ryd\-berg series, but
not the electric field.

\begin{figure}  [h]
\centerline{\ \epsfxsize=3in \epsfbox{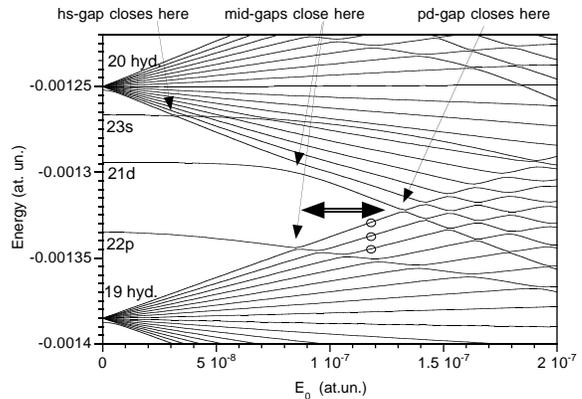}}
\caption
{Stark map of Rb in the vicinity of the $n=19$ and  $n=20$
hydrogenic manifolds.}
\label{starkmap}
\end{figure}

The features of the above type a) 
develop as follows. As $n$ is increased, the 
electric field first manifests itself
in the appearance of $p$-lines and of 
triangular-shaped $h$-features (see Fig.~\ref{spectra1}),
caused by electric-field-induced state mixing.
Due to the linear Stark effect, the hydrogenic states 
fan out over an energy range $3 n^{2} E$ (atomic units) 
(\cite{Gal} and Fig.~\ref{starkmap}).
As $n$ increases further, the $h$-features 
progressively expand and fill in the 
spectral gaps of originally zero oscillator strength 
between the discrete non-hydrogenic $s$-, $p$- and $d$-lines. There are
three types of such spectral gaps, which we label 
$hs$-, $mid$- and $pd$-gaps. 
The gaps become filled in with significant
oscillator strength in exactly that order, at quite
well defined critical wavelengths. Those wavelengths can be
readily converted into critical effective quantum numbers  
$n_{\rm hs}$, $n_{\rm mid}$, and $n_{\rm pd}$.
The $n_{\rm i}$ are robust indicators for the electric field,
because they are solely determined by the
general spreading behavior of the quasi-continuous 
$h$-features. No absolute line strength or 
line-strength ratio needs to be determined.

Assuming a constant electric field
$E_{0}$, the values of
$n_{\rm i}$ at which the spectral gaps 
become filled in can be  related to 
$E_{0}$ using the Stark map 
of Rb. As shown in Fig.~\ref{starkmap}, the $hs$-gap below the
hydrogenic levels with principal quantum number $n$ 
disappears when  the energy shift of the
most red-shifted hydrogenic state, which 
is approximately $-3 n^{2} E_{0} / 2$, equals the splitting $0.13 n^{-3}$
between the unshifted hydrogenic energy and the next-lower 
$s$-state. Further, Fig.~\ref{starkmap} shows that the 
$mid$-gaps, i.e. the gaps between 
the states $(n+3)s$ and $(n+1)d$ and  between the 
field-free hydrogenic states $n-1$ and the states $(n+2)p$,
become entirely covered with
hydrogenic states at practically identical electric fields.  
Finally, the $pd$-gaps disappear when neighboring hydrogenic 
manifolds meet, i.e. when $3 n^{2} E_{0} = n^{-3}$.  
The relations between $E_{0}$ and the critical
quantum numbers $n_{\rm i0}$ are obtained as

\begin{eqnarray}
E_{0} = && 0.086 n_{\rm hs0}^{-5} \nonumber\\
E_{0} = && 0.23 n_{\rm mid0}^{-5} \nonumber\\
E_{0} = && 0.33 n_{\rm pd0}^{-5} \, {\rm (atomic\, units)} .
\label{IT0}
\end{eqnarray}

The zeros in the subscripts indicate that these relations hold for a 
constant electric field. Atomic species with quantum defects different 
from Rb would follow similar relations with different 
numerical factors. Further, the 
Eqs.~\ref{IT0} share, as expected, 
the $n^{-5}$ dependence of the familiar Inglis-Teller
relation \cite{Itrel}.

\begin{figure}  [h]
\centerline{\ \epsfxsize=3.5in \epsfbox{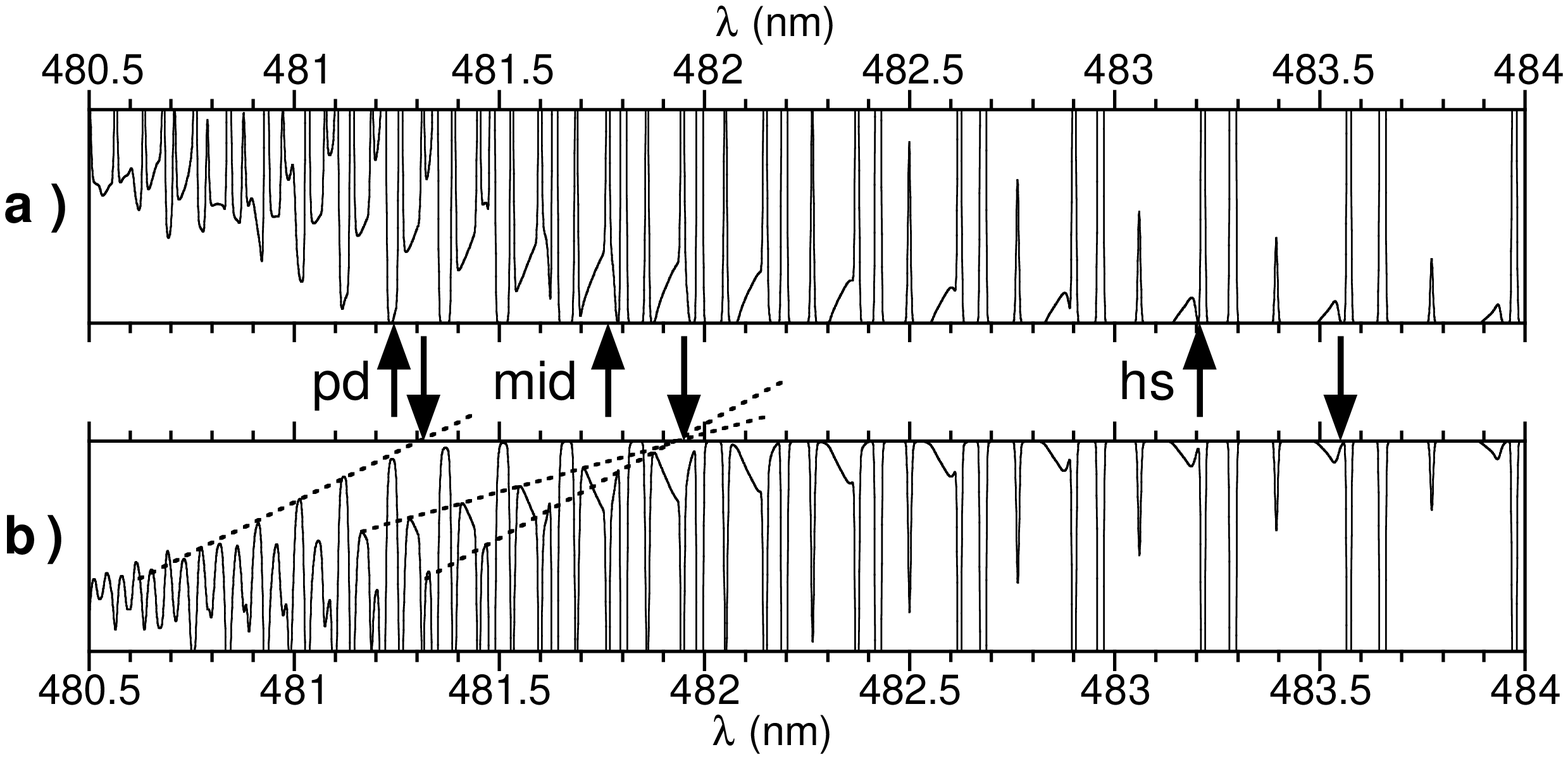}}
\caption
{Calculated isotropic excitation spectra of 
$^{87}$Rb 5P atoms into Ryd\-berg states.
Panel a) is for a homogeneous electric field 
$E_{0} = 6 \times 10^{-9}$at.~un.$= 31$V/cm. Panel
b) shows the excitation spectrum - upside down - in a Gaussian cloud
of $4 \times 10^{5}$ Rb$^{+}$ ions that has a most probable
electric field of $E_{\rm max} = 31$V/cm. The arrows 
mark the critical wavelengths where 
the spectral gaps of low oscillator strength become filled in.}
\label{spectra2}
\end{figure}

To model our experimental spectra, we have 
calculated the photoabsorption spectra $S(E_{0}, \lambda )$ of randomly
oriented  atoms in constant electric fields $E_{0}$. 
Since from the $5P$ intermediate state we excite 
three manifolds of the magnetic quantum number $\vert m \vert $,
one  spectrum $S(E_{0}, \lambda )$ is a properly weighted 
sum of an $m=0$, an $m=1$ and an $m=2$ spectrum.
The spectra are obtained  as 
outlined in \cite{Klep80} using a spherical 
basis  set of all states up to $n=90$.
To account for the laser linewidth, the  spectra are 
convoluted with Gaussians of a few GHz FWHM.  
The calculated spectra, one of which is shown in Fig.~\ref{spectra2} a),
display the same qualitative features 
as our experimental spectra. This applies, in particular, 
to the triangular shape of the $h$-features, which 
is produced by the unresolved excitation of 
the states in the hydrogenic manifolds.

An investigation of the calculated photoexcitation spectra
reveals that the closing of the $pd$-gap appears particularly 
sudden due to the following peculiarity: 
for parameters between the $mid$-gap and the $pd$-gap closing,
indicated by a double-arrow in Fig.~\ref{starkmap}, the 
blue-shifting hydrogenic levels within the $pd$-gap,
marked with open circles  in Fig.~\ref{starkmap},
carry much less oscillator strength
than all the other nearby hydrogenic levels. 
As a result, the whole $pd$-gap appears 
virtually oscillator-strength-free up
until the $pd$-gap closing condition  $3 n^{2} E_{0} \ge n^{-3}$  is 
satisfied.  Once the $pd$-gap closing condition 
is satisfied, the oscillator strength of the
hydrogenic states inside the $pd$-gap increases
rather abruptly. This behavior 
enhances the clarity of the $pd$-gap closing.

\begin{figure}  [h]
\centerline{\ \epsfxsize=2.5in \epsfbox{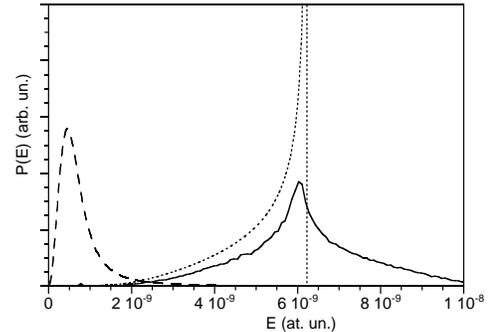}}
\caption
{Electric-field distributions $P(E)$ in Gaussian ion clouds with $N=4 \times 
10^{5}$ and $\sigma =0.2$mm. Solid: Distribution 
obtained by random placement of discrete ion charges
following a Gaussian density distribution. Dotted: 
Field distribution of the corresponding 
{\em continuous} charge density. 
Dashed: Holtsmark distribution of the microfields
in a neutral plasma with the same ion density 
as in the center of the above Gaussian ion cloud.
}
\label{edist}
\end{figure}

The electric field experienced by the 
$5P_{3/2}$-atoms is not homogeneous but
follows a probability  distribution $P(E)$.
Assuming identical Gaussian density profiles 
for the ions and the $5P_{3/2}$-atoms embedded
in between them, the distribution $P(E)$ can 
be calculated by a random placement of ion charges
and  $5P_{3/2}$-atoms using  a radial distribution function
$P(r) =  4 \pi r^{2} (2 \pi \sigma^{2})^{-3/2} 
{\rm exp} ( - r^{2} / (2 \sigma^{2} ))$.  
The resultant field distributions $P(E)$ have well defined 
most probable electric fields $E_{\rm max}$, which
practically coincide  with the singularities of the  
distributions produced by the corresponding
smoothed charge distributions
(compare the solid and dotted curves in  Fig.~\ref{edist}).
The effect of the microfields of the 
discrete ion charges largely is to smear out the 
singularities produced by the smoothed charge distributions,
adding fields significantly larger than $E_{\rm max}$ to the
field distribution. Deviations of the
density profiles of the ions and the $5P_{3/2}$-atoms 
from Gaussians will modify $P(E)$, a dominant peak of 
$P(E)$. However, a most probable field $E_{\rm max}$
and a microfield-induced tail of fields $E>E_{\rm max}$ will
remain.

We have used simulated electric-field
distributions $P(E)$ to form averages of 
the spectra $S(E_{0},\lambda)$
weighted by $P(E)$. As shown in
panel b) of Fig.~\ref{spectra2},
one  can still easily obtain critical quantum numbers
$n_{\rm hs}$, $n_{\rm mid}$, and $n_{\rm  pd}$
at which the respective regions of low oscillator strength 
disappear. The excitation spectra
a) and b) in Fig.~\ref{spectra2} are, in fact,
quite similar, because the $P(E)$ are clearly dominated
by fields near the most probable one. 
The microfield-induced spread of $P(E)$ from $E_{\rm max}$ towards
higher fields causes the spectral gaps 
with low oscillator strength  
to become filled in at values of $n$ that are slightly lower 
than in the case of a homogeneous field $E_{0} = E_{\rm max}$.

Evaluating a number of spectra as shown in
Fig.~\ref{spectra2} b), we have determined the  
relations between the most probable fields $E_{\rm max}$ and
the critical effective quantum numbers $n_{\rm hs}$, $n_{\rm mid}$, 
and $n_{\rm pd}$ for Ryd\-berg atoms excited 
in Gaussian ion clouds with a FWHM of about 1mm:

\begin{eqnarray}
E_{\rm max} = && 0.050 n_{\rm hs}^{-5} \nonumber\\
E_{\rm max} = && 0.145 n_{\rm mid}^{-5} \nonumber\\
E_{\rm max} = && 0.26  n_{\rm pd}^{-5} \, {\rm (atomic\, units)} ,
\label{IT}
\end{eqnarray}

\begin{figure}  [h]
\centerline{\ \epsfxsize=3.5in \epsfbox{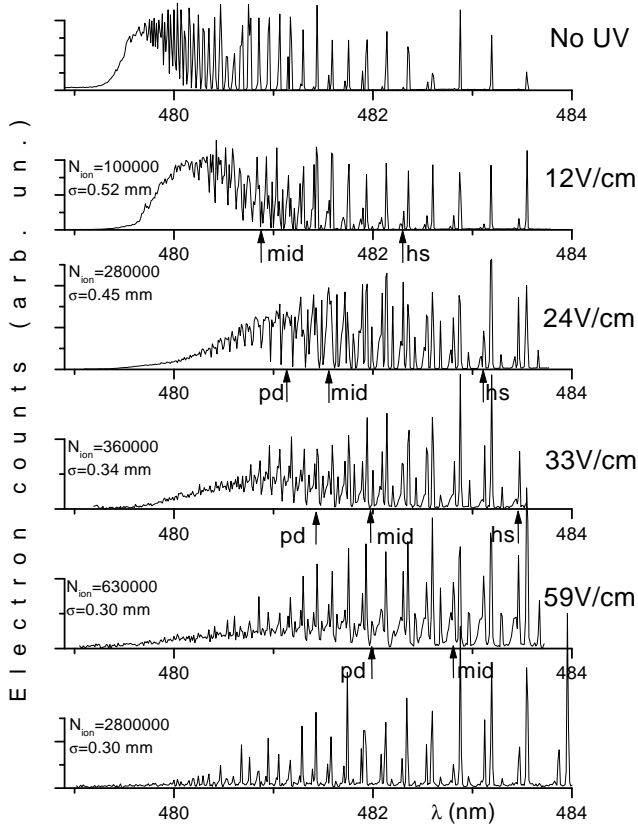}}
\caption
{Experimental excitation spectra of 
$^{87}$Rb 5P$_{3/2}$ atoms embedded in
photo-excited plasmas that contain the indicated
numbers $N_{\rm ion}$ of Rb$^+$-ions.
The $\sigma$-parameters of the 
utilized Gaussian clouds of laser-cooled atoms 
are also shown. The most probable electric 
fields $E_{\rm max}$, shown on the right, 
are spectroscopically 
determined and carry an uncertainty 
of about $20 \%$.}
\label{master}
\end{figure}

We have used Eqs.~\ref{IT} to experimentally 
determine the values 
of $E_{\rm max}$ in ion clouds with 
diameters of $\sim 1$mm and various ion numbers. 
The spectra obtained in a few such experiments 
are displayed in Fig.~\ref{master}.
The critical wavelengths, indicated by arrows,
have been determined as shown in
Figs.~\ref{spectra1} and  \ref{spectra2}, and have
been converted into critical 
quantum numbers $n_{\rm hs}$, $n_{\rm mid}$ 
and $n_{\rm pd}$. Using Eq.~\ref{IT}, up to three
spectroscopic measurements of $E_{\rm max}$ 
are obtained for each spectrum.  
The relative uncertainty of each single electric-field value is
$30 \%$. Within that uncertainty, 
the different electric-field values obtained for 
the individual spectra agree well.   
The electric field values quoted in Fig.~\ref{master}
are the averages over all field values obtained for the
respective spectra.

The most probable field $E_{\rm max}$ in Gaussian ion clouds
can be calculated as 
$E_{\rm max}= ( 3.1 \times 10^{-10} N_{\rm ion} \sigma^{-2})$Vm, 
with the RMS cloud size $\sigma$ and ion number $N_{\rm ion}$.  
We have independently 
deduced $N_{\rm ion}$ from measurements of the MOT fluorescence.
The values of $\sigma$ could be determined from cuts through CCD images 
of the atom cloud fluorescence, yielding
values of $\sigma$ ranging from 0.3mm to 0.5mm.
For large atom clouds, we have found that 
the measured fields exceed the calculated ones
by up to a factor of ten. As the atom cloud size is 
decreased, the discrepancy drops to about a factor of two.
The fields deduced from the spectra are, of course, the 
ones to be trusted. The most likely explanation
for the discrepancy and its 
dependence on the size of the 
atomic cloud is that the ions were actually not excited
in the whole atomic cloud, but were concentrated 
in a significantly smaller volume
determined by the diameters of the excitation
laser pulses. 

If the number of ions in a Gaussian  cloud
exceeds $7.8 \times 10^{5} {\rm mm}^{-1}  \sigma$,
the potential depth $U_{0}$ of the ion plasma exceeds $0.9$eV, and
a fraction of the UV-excited photoelectrons will be retained 
\cite{Rol99}. This is clearly
 the case in the lowest two spectra of Fig.~\ref{master},
where we exceed the critical ion number by factors  $>2.5$ and
$> 10$, respectively. If the fraction of 
retained electrons is large, a
macroscopically close-to-neutral plasma core
is formed, which is surrounded by a positively charged ionic shell.
The electric field in the core is close to zero,
while the shell 
carries a large electric field.
The distribution $P(E)$ will thus become bi-modal, featuring
a peak near $E=0$ and a high-field peak.  
Ryd\-berg atoms excited in the shell
produce the spectroscopic field signatures discussed in 
this paper, while atoms excited in the core
contribute a low-field excitation spectrum with resolved 
Ryd\-berg resonances up to large $n$.
The excitation spectra shown in the two lowest 
plots of Fig.~\ref{master} appear, in fact, as superpositions 
of a high-field and a low-field spectrum, as expected for bi-modal 
distributions $P(E)$. In the lowest curve, the high-field 
component of $P(E)$ apparently 
is strongly dominated by the low-field one, making it hard to determine 
the value of the electric field in the shell. The presence of a strong 
electric field, can, however, be guessed from 
the appearance of $p$-lines at 
long wavelengths.

We note that in the case of perfect macroscopic neutrality of the core
the microfields in the core would follow a Holtsmark 
distribution \cite{Holts}, as displayed in  Fig.~\ref{edist}.
For our charge carrier densities and laser linewidths,
the Holtsmark fields are too low to be detectable. 
 
In this work, we have used
spectroscopic tools to probe the 
electric fields in 
small, laser-excited plasmas. The technique has a large future 
potential: using crossed and focused Ryd\-berg atom
excitation beams, it will be possible to 
perform spatially resolved 
measurements of the plasma electric fields.  
A variation of the time delay between the
UV pulse and the blue probe pulse will allow 
one to study the plasma expansion via
the decay of the electric field.
The use of a laser with a narrow bandwidth
($\sim 1$MHz) will allow one to 
detect smaller fields, such as the Holtsmark fields, 
and to make precise measurements of the 
detailed electric-field 
distribution $P(E)$. Measurements of $P(E)$
could reveal inhomogeneities and  
instabilities in the plasma, which are expected
to produce structures in $P(E)$
at excessive fields.  Spatial order in the charge carrier arrangement,
such as Wigner crystals  \cite{order}  and 
the formation of crystalline Ryd\-berg matter \cite{RM}
may become detectable via measurements of $P(E)$.

Support by NSF is acknowledged. 
We thank Prof. P. Bucksbaum for inspiring discussions and 
generous loaning of equipment.

\end{document}